\documentclass[aps,twocolumn,superscriptaddress,prb,showpacs,amsmath,amssymb]{revtex4}% Physical Review B

\usepackage{graphicx}% Include figure files
\usepackage{dcolumn}% Align table columns on decimal point
\usepackage{bm}% bold math
\usepackage[english]{babel}

\begin{document}

%\preprint{APS/123-QED}

\title{Manipulating  the exciton fine structure of single CdTe/ZnTe quantum dots by an in-plane magnetic field}% Force line breaks with \\

\author{ K.~Kowalik}\email[Corresponding author
:]{kowalik@lpn.cnrs.fr} \affiliation{Institute of Experimental Physics, Warsaw University, Ho\.za 69, 00-681 Warszawa,
Poland} \affiliation{Laboratoire de Photonique et Nanostructures-CNRS, Route de Nozay, 91460 Marcoussis, France}
\author{ O.~Krebs}
\affiliation{Laboratoire de Photonique et Nanostructures-CNRS, Route de Nozay, 91460 Marcoussis, France}
\author{ A.~Golnik}
\author{ J.~Suffczy\'nski}
\affiliation{Institute of Experimental Physics, Warsaw University, Ho\.za 69, 00-681 Warszawa, Poland}
\author{ P.~Wojnar}
\author{ J.~Kossut}
\affiliation{Institute of Physics, Polish Academy of Sciences, Al. Lotnik\'ow 32/46, 02-668 Warszawa, Poland}
\author{ J.~A.~Gaj} \affiliation{Institute of Experimental Physics, Warsaw University, Ho\.za 69, 00-681
Warszawa, Poland}
\author{ P.~Voisin}
\affiliation{Laboratoire de Photonique et Nanostructures-CNRS, Route de Nozay, 91460 Marcoussis, France}

\date{\today}% It is always \today, today,
             %  but any date may be explicitly specified

\begin{abstract}
Polarization resolved photoluminescence (PL) spectroscopy of individual CdTe/ZnTe quantum dots is investigated in the presence of external
in-plane magnetic field. We find that the excitonic fine structure strongly depends on the magnitude and direction of applied field. The
splitting between "bright" and "dark" states increases with the magnetic field, whereas the anisotropic exchange splitting of the bright
excitons can be reduced or enhanced, depending on the field direction. Increase (decrease) is observed when the field is applied parallel to the
PL polarization direction of the lower (upper) energy exciton. For intermediate fields, we observe a  rotation of the PL polarization
orientation. The results are discussed in terms of an effective spin Hamiltonian derived for the exciton ground state.
\end{abstract}

\pacs{71.35.Ji,71.70.Gm,73.21.La,75.75.+a}% PACS, the
%Physics and Astronomy
% Classification Scheme.
%\keywords{Suggested keywords}%Use showkeys class option if keyword
%display desirede

\maketitle

\section{Introduction}

%\begin{introduction}
Semiconductor quantum dots (QDs) are unique non-classical light emitters. In addition to their now well-known single photon emission properties,
their potential as sources of entangled photons on demand \cite{Akopian, Shields, Shields2} was recently demonstrated. The main obstacle to
polarization entanglement of photons emitted in the biexciton-exciton radiative cascade is the lifted of degeneracy of the two optically active
exciton states due to the anisotropic electron-hole exchange interaction. It is revealed in experiment by a splitting of the exciton and
biexciton lines into doublets with orthogonal linear polarizations. The origin of the symmetry breakdown governing this fine structure splitting
(FSS)  is not fully established. FSS can result from any combination of in-plane shape anisotropy of a dot (elongation of the dot due to
preferential growth direction), piezoelectric potential in the dot vicinity (due to the vertically asymmetric strain field), and local symmetry
breakdown due to chemical bond alignment at the dot interfaces~\cite{Zunger, Sequin}. As controlling FSS is of utmost importance for quantum
optics applications, different strategies for restoring higher symmetry were tested, either by influencing material properties of
heterostructures (annealing or strain engineering~\cite{Young, Tartakovskii}) or by applying external perturbations compensating the native
asymmetry : in-plane electric field \cite{Kowalik-APL}, uniaxial strain~\cite{Seidl}, and in-plane magnetic field~\cite{Stevenson, Shields} were
tried. The last-mentioned method has given the most satisfactory results so far in GaAs-based self-assembled QDs. II-VI systems have promising
features in this context, based on their more robust excitonic states allowing to study non-classical light emission at higher
temperatures~\cite{Tinjod-temp}. However, they generally exhibit stronger anisotropy splittings~\cite{Puls}, so it is essential to devise
efficient methods of symmetry control suited to II-VI QDs. In this paper we report a study of exciton FSS in CdTe/ZnTe QDs in the presence
of an in-plane magnetic field and show that it can be increased or decreased, depending on the in-plane  field direction.\\
\section{Sample and Experiment}
%\begin{sample and experiment}
\indent The studied sample was grown by MBE on a $($001$)$-oriented GaAs substrate. It consists of following layers: a thick (4.3$\mu$m) CdTe
buffer, followed by a 0.35$\mu$m ZnTe barrier, a CdTe quantum dot layer, and a 114nm ZnTe cap. QD formation was induced by desorption of
amorphous Tellurium deposited on six monolayers of CdTe \cite{Tinjod}. Before the cap deposition the dot layer was annealed in-situ at
$480^{o}$C for 25 minutes. The sample emission is characterized by a broad micro-luminescence ($\mu$PL) spectrum, with well resolved individual
QD lines appearing on the low energy tail. For the experiment the sample was mounted directly on a specially designed Cassegrain-type microscope
objective immersed in liquid helium \cite{Jasny}, in a cryostat with superconducting coil allowing to apply magnetic field up to 7T. The field
was applied parallel to one of the $\langle110\rangle$ crystallographic axes, in Voigt configuration. For the excitation, a CW doubled YAG laser
at 532nm was focused on a $1-2\mu m^{2}$ area spot on the sample surface. The photoluminescence signal was collected by the same objective,
analyzed using a linear polarizer, filtered by a monochromator, and recorded by a CCD camera. Excellent mechanical stability of this set-up
enabled PL measurement of a single QD for many hours. For well isolated dots, lineshape fits allowed us to determine the line position with a
precision of 30$\mu$eV. A distinctive characteristic of CdTe quantum dots is that their optical axes are randomly oriented
\cite{Kudelski-ICPS,Marsal}, contrarily to InAs QDs where they are clamped to the $\langle110\rangle$ direction. This allows simultaneous
measurements of different relative orientations of dot and applied field in a fixed geometry, by selecting different dots and rotating the
analyzer accordingly. We define the dot orientation as the polarization direction of
the lower component of the excitonic doublet and denote $\theta$ the angle between the field and this direction.\\
\section{Results}
%\begin{results}
\begin{figure}[h]
\includegraphics[width=0.45 \textwidth,keepaspectratio]{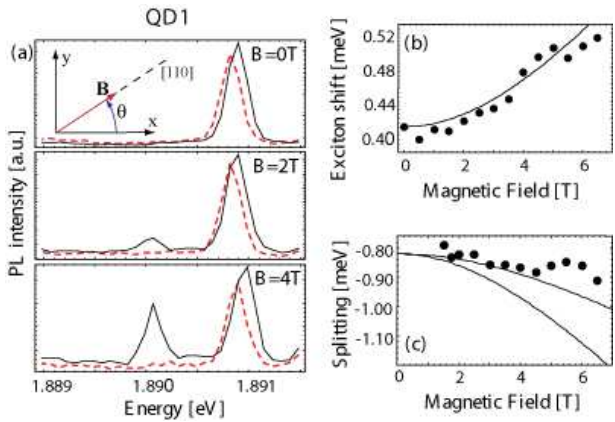}%
\caption{(Color online)(a) Linearly polarized excitonic doublet (solid and dashed lines) for magnetic fields $B=$0,~2, and 4~T in a dot
characterized by $\theta =20^{\circ}$. Inset shows the direction of the magnetic field with respect to the QD polarization eigenaxes.(b) Mean
energy of the "bright" excitonic doublet versus magnetic field and (c) splitting of bright to dark states vs magnetic field. For (b) and (c)
Lines are theoretical curves calculated with the model described in the text.}\label{fig1}
\end{figure}
\begin{figure}[h]
\includegraphics[width=0.45 \textwidth,keepaspectratio]{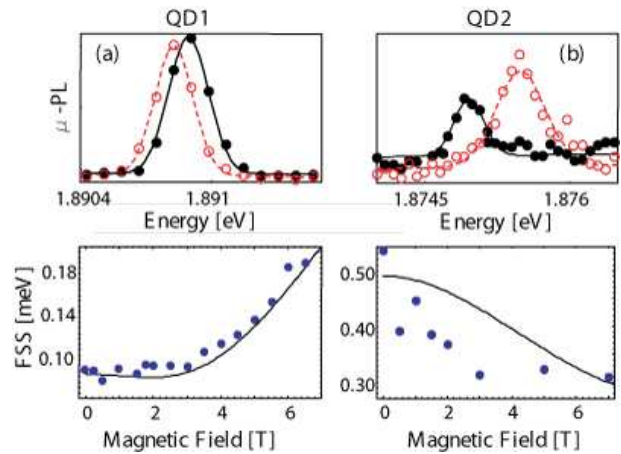}%
\caption{(Color online) Upper panel: $\mu$-PL spectra at $B=$0~T for two dots ((a) and (b)) with perpendicular anisotropy orientations,
$\theta$=20$^{\circ}$ and $\theta$=110$^{\circ}$, (open symbols and dashed line: measured at $20^{\circ}$, closed symbols and solid line:
measured at $110^{\circ}$). Lower panel: Fine structure splitting (FSS) vs in-plane magnetic field $B$ for the dots from upper panel. Solid
lines are theoretical curves according to the model discussed in the last section.}\label{fig2}
\end{figure}

\indent The polarization-resolved $\mu$-PL spectra of a QD at various fields are shown in Fig.~\ref{fig1} for the peculiar case of a dot nearly
oriented along $\bm{B}$ ($\theta=20^{\circ}$). The most salient feature is a line appearing $\sim$1~meV below the excitonic doublet, and
developing when the field is increased. This line is attributed to the dark exciton that becomes optically active due to field-induced mixing of
bright and dark states~\cite{Bayer-PRB65}. Detailed analysis shows in general a blueshift of the excitonic doublet (105~$\mu$eV at 7~T), a
similar increase of the bright-dark exciton splitting, and in this case, an increase of FSS from 87~$\mu$eV at $B=$ 0 to 185~$\mu$eV at $B=$7~T.
In fact, the magnetic field influence on the PL fine structure depends not only on the field magnitude, but also on some  intrinsic properties
and orientation of the quantum dot. In particular an increase of FSS is observed when the field is applied parallel to the orientation of the
lower energy component of excitonic doublet ($\theta\approx0$, see Fig.~\ref{fig2}(a)), while a decrease is produced for perpendicular direction
of the field ($\theta\approx\pi/2$, see Fig.\ref{fig2}(b)).
 \\
\begin{figure}[h]
\begin{center}
\includegraphics[width=0.45 \textwidth,keepaspectratio]{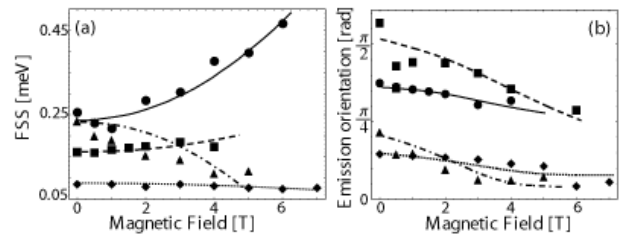}%
\end{center}
\caption{(a) Fine structure splitting (FSS), and (b) PL polarization orientation $\theta$ against in-plane magnetic field of four different QDs
(different symbols). Lines are guide to the eyes.}\label{fig3}
\end{figure}\\
\indent Finally, we have also investigated some dots with a strongly tilted field configuration (i.e. $\theta\approx\pi/4$ or $\theta\approx
3\pi/4$). In such cases, in addition to the FSS change, the dot  orientation shows a clear rotation when increasing the field. Figure~\ref{fig3}
illustrates this effect for only a few selected dots, but observations were the same for all the QDs that we have studied (around 20). As shown,
the reference QD eigenaxis rotates systematically towards the direction of the applied field. This indicates that the in-plane magnetic field
$\bm{B}$ contributes to the FSS by an effective spin splitting of the bright  exciton states characterized by the low (high) energy component
polarized parallel (orthogonal) to the field. This conclusion is also supported by the corresponding FSS modification which depends both
qualitatively  and quantitatively   on the initial angle $\theta$ between the field direction and QD orientation. The initial angle has to be
close to $\pi/2$ in order to get a reduction of FSS. If not, there is first a rotation of the optical orientation followed by an increase of FSS
due to the field. The theoretical discussion presented below sheds some light on this qualitative description.\\
\section{Discussion}
\begin{figure}[h]
\begin{center}
\includegraphics[width=0.45 \textwidth,keepaspectratio]{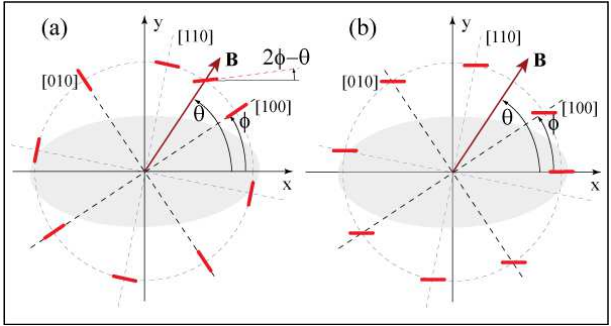}%
\end{center}
\caption{(Color online) Schematics of the crystallographic axis configuration with respect to the QD principal axes $x,\,y$. Thick red dashes
represent the exciton polarization orientation  produced by the field-induced splitting only for different directions of the  field. (a) Effect
of the cubic term (see text), (b) effect of the linear term, including heavy-hole to light-hole coupling. The grey shaded area represents a  QD
elongated along the polarization eigenaxis $x$.} \label{schema}
\end{figure}

\begin{figure}[h]
\begin{center}
\includegraphics[width=0.45 \textwidth,keepaspectratio]{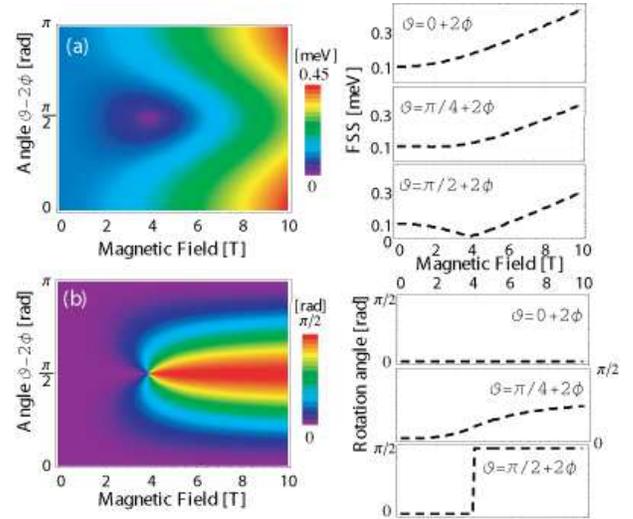}%
\end{center}
\caption{(a) FSS absolute value vs in-plane field direction $\theta-2 \phi$ and magnitude encoded on a color scale. Cross-sections for three
directions of the field are displayed on the right-hand side. (b)Rotation angle $\Delta\theta$ of the PL polarization orientation vs in-plane
magnetic field direction $\theta$ and magnitude. Three cross-sections for $\theta=0,\pi/4$~and~$\pi/2$ are also shown. Calculations are made
with an isotropic $g$~factor ($\beta$=0), $\delta_{0}$=~80~$\mu$eV  and a FSS $\delta_{1}$=~80~$\mu$eV in zero field.}\label{fig_q-term}
\end{figure}

\indent The  electron-hole exchange Hamiltonian responsible for the ground state exciton fine structure in an anisotropic quantum dot can be
represented by~\cite{Ivchenko} :
\begin{equation}
\hat{H}_{ex}=\frac{\delta_{0}}{2}\hat{\sigma}^{e}_{z}\hat{\sigma}^{h}_{z}+
\frac{\delta_{1}}{4}(\hat{\sigma}^{e}_{x}\hat{\sigma}^{h}_{x}-\hat{\sigma}^{e}_{y}\hat{\sigma}^{h}_{y})+
\frac{\delta_{2}}{4}(\hat{\sigma}^{e}_{x}\hat{\sigma}^{h}_{x}+\hat{\sigma}^{e}_{y}\hat{\sigma}^{h}_{y})
\label{Exchange}
\end{equation}
where the Pauli matrices $\sigma_{i}^{e,h}$ act on the spin components of the electron (e) or hole (h) respectively. Here, we used a $\pm 1/2$
pseudo-spin  to describe the QD hole ground states with angular momentum $J_{z}=\mp 3/2$ along $z$. The quantities $\delta_{0}$, $\delta_{1}$,
and $\delta_{2}$ describe the exciton quartet fine structure as follows : $\delta_{0}$ \-- between states of  angular momentum $|M|=1 $ and
$|M|=2$ (or $\sigma_z^{e}+\sigma_z^{h}$=0), $|\delta_{1}|$ (i.e. FSS) \-- between the components of the optically active doublet ($ M=\pm 1$),
and $|\delta_{2}|$ \-- between the dark states ($ M=\pm 2$). These parameters are determined by the quantum dot properties (size, shape,
composition, strain field, etc). In this formalism,  the  arbitrary $x$,~$y$ directions of the Pauli matrices correspond to the eigenaxes of the
QD. In the following we assume that the parameters $\delta_{0}$, $\delta_{1}$, and $\delta_{2}$ are not directly modified by the in-plane
magnetic field, although for high field values the magnetic confinement likely affects the electron-hole exchange. Therefore, to the first order
we only consider   the Zeeman Hamiltonian to describe the effect of the in-plane field $\bm{B}_{\perp}$ as recently done for self-assembled InAs
QDs~\cite{Stevenson}. To derive properly the $g$~factor for hole ground states, we start from the general expression available for bulk excitons
and given by~\cite{van-Kesteren}:
\begin{equation}
 \hat{H}_{Z}^{bulk}(\bm{B})=\mu_{B}\sum
_{i=x,y,z}(\frac{1}{2}g^{e}\hat{\sigma}_{i}^{e}B_{i}-2\kappa\hat{J}_{i}B_{i}-2q\hat{J}^{3}_{i}B_{i}) \label{Bulk_HZ}
\end{equation}
where $\mu_{B}$ is the Bohr magneton, $g^{e}$ is the electron Land\'{e} factor, $q, \kappa$ are Luttinger coefficients and the $\hat{J}_{i}$'s
are   the  angular momentum projections of the Bloch states in the $\Gamma_{8}$ hole band along the crystallographic axes $\langle100\rangle$.
Usually, the main term driving the hole Zeeman splitting is the linear term $-2 \mu_{B}\kappa\hat{\bm{J}}\cdot\bm{B}$ while the cubic term in
Eq.~\ref{Bulk_HZ} is considered as negligible. Yet, for a transverse magnetic field, only the hole states which differ by $|\Delta J|$=1 are
coupled by $\hat{J}_{x}$ or $\hat{J}_{y}$. As a result, in the quantum dots investigated here, the hole ground states which are essentially pure
heavy-holes with $J_{z}=\pm3/2$ are not directly split by this term~\cite{AngularMomentum}. To obtain a non-zero transverse $g$~factor, which is
required to modify the exciton FSS~\cite{Stevenson}, it seems thus necessary either to take into account the cubic term in Eq.~\ref{Bulk_HZ}, or
to include in the model a light-hole doublet state ($J_{z}=\pm1/2$) split by an energy $\Delta_{h-l}$ of a few tens meV's from the hole ground
state doublet. Actually, the sole Zeeman coupling to the light-hole states produces  only a weak third-order contribution to the heavy hole
splitting in a transverse magnetic field. We could conclude that only the cubic term contributes to the effective $g$~factor. However, including
the light-holes also enables us to take into account the QD symmetry reduction to $C_{2v}$ or even $C_{2}$ (responsible for the FSS) which
implies a direct coupling between the heavy and light hole ground states. The latter is proportional to the symmetrized product of the in-plane
angular momentum  components $\{\hat{J}_{x'}\hat{J}_{y'}\}$~\cite{Ivchenko,PRB63-Toropov}. Here, the indexes $x',\, y'$ denote axes which are
rotated by $\pi/4$ with respect to the QD eigenaxes. After a $-\pi/4$ rotation  to use the same referential axes as Eq.~(\ref{Exchange}), we
obtain the following Hamiltonian for the heavy-hole to light-hole coupling:
\begin{equation}
 \hat{H}_{h\!-\!l}=\beta \left( \hat{J}_{x}^{2}-\hat{J}_{y}^{2}\right)
\label{C2v_term}
\end{equation}
where $\beta$ represents  the strength of the coupling. In the above formalism, based essentially on symmetry considerations, the respective
signs of $\delta_{1}$ and $\beta$ are not \textsl{a priori} correlated although they are necessarily determined by the features of a given QD.
This unknown sign correlation could  however reveal of importance for the control of the FSS as discussed below~\cite{HexcC2v} and  somehow
enlightens the concept of \textquotedblleft inverted\textquotedblright  FSS in Ref.~\onlinecite{Stevenson}. Experimentally the perturbation
$\hat{H}_{h\!-\!l}$ leads also to dichroism of the ground state excitonic transition (i.e. a difference in oscillator strength of the
linearly-polarized doublet components) as reported in the past for the quantum wells of $C_{2v}$ symmetry~\cite{PRB63-Toropov} and more recently
for trions in CdSe QDs~\cite{Koudinov}. It is worth mentioning that in InAs QD's similar dichroism has been reported and that no correlation was
found between the sign of the linear polarization degree (related to $\beta$) and the sign of $\delta_{1}$~\cite{APL-Favero}. For taking into
account both $\hat{H}_{Z}^{bulk}$ and $\hat{H}_{h\!-\!l}$ we have to introduce the angle $\phi$ between the QD eigenaxis $x$ and the
crystallographic direction [100] in order to rotate the cubic term of $\hat{H}_{Z}^{bulk}$ in the QD coordinate frame (see Fig.~\ref{schema}).
In this way, we obtain the effective Zeeman Hamiltonian  to the first order in $\beta/\Delta_{h\!-\!l}$ and in the basis of electron spin and
hole pseudo-spin :
\begin{equation}
 \hat{H}_{Z}(\bm{B}_{\perp})=\frac{1}{2}\mu_{B}\left(\sum_{i}g^{e}\hat{\sigma}^{e}_{i}B_{i}+\sum_{i,j}
\hat{\sigma}^{h}_{i}g^{h}_{ij}B_{j}\right)\label{QD_HZ}
\end{equation}
with the hole $g$~factor tensor :
\begin{eqnarray}
g^{h}&=&3q\left(
  \begin{array}{cc}
    \cos4\phi +\rho_{g} & \sin4\phi \\
    -\sin4\phi & \cos4\phi -\rho_{g}\\
  \end{array}
\right)\label{g_factor}\\
\nonumber\mbox{where}\quad \rho_{g}&=&\frac{(4\kappa+7q)\beta}{q\Delta_{h\!-\! l}}
\end{eqnarray}
 As it could be expected, the transverse hole $g$~factor gets anisotropic due to the term proportional to $\beta$ as already emphasized in
Ref.~\onlinecite{Koudinov}. Of course, the parameters $g^{e}$,~$\kappa$ and $q$ in Eqs.~\ref{QD_HZ},~\ref{g_factor} are not the bulk parameters
of a quantum dot (or a barrier) material: in QDs the strong confinement of  eigenstates considerably affects the values of these parameters as
predicted~\cite{PhysRevB.58.16353,PRL96-Pryor} and experimentally observed~\cite{Bayer-PRB65}. In particular the mixing allowed in $D_{2d}$
symmetry  between heavy-hole and light-hole with \emph{envelope} wave functions of different angular momentum may explain the significative
value often reported for the transverse hole $g$~factor~\cite{AngularMomentum}. In the following we assume first a symmetric transverse
$g$~factor for the hole ($\beta$=0) and then discuss the effect produced by an  antisymmetric term ($\beta\neq$0).
\\
\indent The total Hamiltonian $\hat{H}_{ex}+\hat{H}_{Z}$ enables us to predict the evolution of the bright state FSS as a function of
$\bm{B}_{\perp}$. The results are displayed in Fig.~\ref{fig_q-term} which shows the absolute splitting of the bright excitons  (panel (a))  and
their polarization rotation angle $\Delta\theta$ (panel (b)). Both are plotted as a function of field magnitude and orientation represented here
by $\theta-2\phi$.  Before commenting further these figures, let us consider the effect of the cubic term in $\hat{H}_{Z}$ for a symmetrical QD
with $\delta_{1}=0$ (and $\beta=0$). In this case, the choice of the QD $x,\, y$ axes to define the angles $\theta$ and $\phi$ is arbitrary.
Choosing $\phi=0$ shows that the  hole $g$~factor tensor reduces to the scalar value $3q$ which leads to an isotropic FSS induced by the field
when its orientation is varied. On the other hand, if we fix $\theta=0$ we observe that in the referential attached to the field, the
polarization of the upper excitonic doublet split by the field rotates faster by an angle $2\phi$ than the field  from the [100] axis. This is
illustrated in the left-hand part of Fig.~\ref{schema}. In the general case, it is thus clear that depending on the field direction $\theta$,
the field-induced splitting will add to or subtract from an initial finite splitting $\delta_{1}$. Calculations presented in
Fig.~\ref{fig_q-term}~(a) show that cancelation can  be achieved for $\theta=\pi/2 + 2\phi$ (changing the sign of $q$ with respect to $g^{e}$
shifts this angle by $\pm\pi/2$), i.e. for a field oriented symmetrically to the low energy component of the bright doublet ($y$ axis) with
respect to the crystallographic direction [100]. A small discrepancy of the field orientation leads to a continuous rotation of the eigenaxes
directions when passing near the critical point $(B_{crit.},\pi/2+ 2\phi)$ of exact cancelation as shown in Fig.~\ref{fig_q-term}(b).
Nevertheless, choosing correctly the field direction with respect to the QD orientation allows in principle to reduce the FSS of any QDs. This
point was not clearly established in the analytical treatment presented by Stevenson~\textit{et al.}~\cite{Stevenson} where however only the
isotropic contribution of Eq.~(\ref{QD_HZ}) was considered~\cite{InAsQDs}.\\

\begin{figure}[h]
\begin{center}
\includegraphics[width=0.45 \textwidth,keepaspectratio]{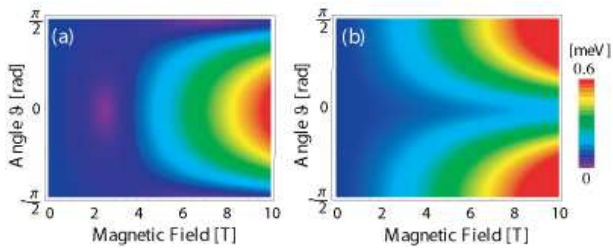}%
\end{center}
\caption{FSS absolute value  on a color scale vs in-plane field direction $\theta-2 \phi$ and magnitude for a strong $g$~factor anisotropy
$\rho_{g}$. Depending on its sign with respect to the initial splitting (here $\delta_{1}$=~80$\mu$eV) it produces very different behaviors :
(a) $\rho_{g}$=-1.3, (b) $\rho_{g}$=+1.3. We used the same average $g$~factor $\bar{g}_{h}$=0.78, $g^{e}$=1.75 and $\delta_{0}$=~0.8~meV  as in
Fig.~\ref{fig_q-term}.}\label{fig_Beta-term}
\end{figure}

\indent Taking now into account the  $g$~factor anisotropy ($\rho_{g}\propto\beta\neq$0) may considerably change the above phenomenology. As
shown in Fig.~\ref{fig_Beta-term} the key feature turns out to be the sign of $\rho_{g}$ (determined by that of $\kappa\beta$) with respect to
$\delta_{1}$. If opposite, we find that it is still possible to reduce to zero the FSS for $\theta=\pi/2+2\phi$, and actually when the
antisymmetric part really dominates ($|\rho_{g}|\gg$1) this can be achieved for any field direction. On the contrary, for $\rho_{g}$ and
$\delta_{1}$ of same sign  the critical field $B_{crit.}$ for which cancelation could be achieved, diverges when $|\rho_{g}|$ approaches 1, a
situation which indeed corresponds to  $g^{h}_{yy}$=0.  For larger values of the anisotropy the magnetic field produces an increase of the FSS
whatever  its in-plane orientation $\theta$ is, as illustrated in Fig.~\ref{fig_Beta-term}(b). In both cases, when the anisotropy dominates
($|\rho_{g}|\gg1$) the principal axes of the PL polarization remain essentially parallel to their initial orientation (see
Fig.~\ref{schema}(b)), in contrast to the case  $\beta=0$ as obvious   in Fig.~\ref{fig_q-term}(b) for fields above $\sim$4T. Our analysis
reveals thus that in the case of a strong anisotropy of the hole $g$~factor the possibility of tuning the QD FSS by a magnetic field depends
essentially on the sign of  $\beta$ with respect to $\delta_1$.

 \indent In our experiments we did not vary the in-plane field orientation $\theta$ for a given QD. Therefore it turns out to be rather delicate
to determine the strength of the in-plane $g$~factor anisotropy. Yet, the fact that  in most cases we observed a rotation of the QD principal
axes towards those defined by the field indicates that the average value ($3q$) of the $g$~factor likely dominates over its anisotropic
contribution $\propto \beta$. Besides, we also did not observe significative degree of linear polarization between the PL intensities of both
exciton components, which means that $\beta/\Delta_{h\!-\!l}\ll 1$. This situation clearly differs from that reported for CdSe
quantum dots~\cite{Koudinov}.\\

\indent  For the quantum dot QD1 shown in Figs.~\ref{fig1},~\ref{fig2}, we could produce a good fit of the FSS evolution by taking the values of
$\delta_{1}$ ($\delta_{0}$) observed (extrapolated) in zero magnetic field, and $\delta_{2}$ assumed to be around 1~$\mu$eV. The only fitting
parameters were thus the electron and hole $g$~factors considered as isotropic.  We included in the model the field orientation $\theta$  with
respect to the dot main axis, which also defines $\phi$, as in our experiment the field was parallel to one of the cleaved edge of the sample
corresponding to  $|\theta\!-\!\phi|=\pi/4$. We obtained the following $g$~factor values $|g^{e}|=1.75\pm0.1$ and $|g^{h}|=0.78\pm0.1$ for the
fit shown in Fig.~\ref{fig1}. The decrease of FSS observed for QD2 (characterized by $\theta\approx\pi/2$) could also be reproduced by the model
with values $|g^{e}|=1.75\pm0.2$ and $|g^{h}|=0.9\pm0.2$. The dark states being absent from the spectra, $\delta_{0}$=1~meV was arbitrary chosen
for the latter fitting. In the general case of tilted QD orientation the almost systematic rotation of the PL polarization towards the field
direction ($\theta+\Delta\theta\rightarrow 0$ in Fig.~\ref{fig3}(b)) agrees  well with our model with a negligible $g$~factor anisotropy. But
clearly, further experimental investigations consisting in a full mapping of the influence of the magnetic field on FSS as a function of its
in-plane orientation and magnitude should be performed to determine more quantitatively the  $g$~factor anisotropy in these quantum dots.

\indent In all studied QDs, the observed dark states were fully linearly polarized (see for example Fig.~\ref{fig1}(a)) and we noticed only one
component of dark excitonic doublet. The lack of one "dark" state in the spectrum can be explained within our interpretation. For the used
experimental configuration $\phi=\pi/4$, and for quantum dots with eigenaxes parallel to the crystallographic direction $[110]$ or $[-110]$
($\theta=0$ or $=\pi/2$) we obtain the Hamiltonian in a particularly simple form. It may be used to describe the case of QD1, for which $\theta$
is close to zero. After writing down the discussed Hamiltonian in a basis of linearly polarized states~\cite{LinearBasis} it has a following
form: \small
\begin{align}
&\hat{H}_{QD1}=\nonumber\\\frac{1}{2}&\left(
\begin{array}{cccc}
  \delta_{0}+\delta_{1} & 0 & \mu_{B}B (g^{e}+g^{h}) & 0  \\
  0 & \delta_{0}-\delta_{1} & 0 & \mu_{B}B (g^{e}-g^{h})  \\
  \mu_{B}B (g^{e}+g^{h}) & 0 & -\delta_{0}+\delta_{2} & 0 \\
  0 & \mu_{B}B (g^{e}-g^{h})& 0 & -\delta_{0}-\delta_{2}  \\
\end{array}
\right)
\end{align}
\normalsize

The diagonal elements refer to energy levels of states in absence of magnetic field, the non-diagonal ones show the mixing induced by the field.
In this representation it is clearly visible that mixing of dark and bright states occurs independently for each linear polarization. The mixing
matrix elements for $x$ and $y$ polarizations are proportional either to $(g^{e}+g^{h})$ or to $(g^{e}-g^{h})$, respectively. The corresponding
intensities of the transitions are proportional to the squares of corresponding matrix elements. For the obtained values of g-factors the
theoretically predicted intensity ratio for dark transitions is $7.84$. It explains qualitatively the strong asymmetry in the intensities of the
two components of the dark excitonic doublet.

\section{Conclusion}
%\begin{conclusions}
\indent In summary, we have shown that an in-plane magnetic field modifies the fine structure splitting of the excitonic emission of CdTe/ZnTe
quantum dots. This effect depends on the field direction. If applied along one of the main axes of the dot, the field can either increase or
decrease the splitting. If not, a rotation of the bright exciton eigenaxes towards the  axis defined by the field direction is generally
observed together with a change of the splitting. These effects are in qualitative (polarization rotation) and rough quantitative (splitting
variation) agreement with a simple model based on a Zeeman spin Hamiltonian. We find that in the QDs investigated here the anisotropy of the
$g$~factor is likely negligible, in contrast to results reported for other types of self-assembled QDs. This could be an advantage as in this
case the possibility to cancel the fine structure splitting does not seem to be hindered by light---heavy hole mixing.
\\
%\end{conclusions}
\begin{acknowledgments}
This work has been partially supported by Polish Ministry of Science and Higher Education (Grants 1PO3B-114-30 and 2PO3B-015-25). One of us
(K.K.) is supported by the European network of excellence SANDIE. We would like to thank Pr. E. Ivchenko for enlightening discussions.\\
\end{acknowledgments}
\bibliographystyle{apsrev}
%\bibliography{Biblio_Voigt}% Produces the bibliography via BibTeX.

\begin{thebibliography}{27}
\bibitem[{Ang()}]{AngularMomentum}
\bibinfo{note}{For a quantum dot with $D_{2d}$ symmetry, the projection of the
  total angular momentum $\hat{M}_{z}=\hat{L}_{z}^{envelope}+\hat{J}_{z}$ is a
  good quantum number and the hole ground states correspond to $M_{z}=\pm 3/2$.
  Yet, due to the coupling of heavy-hole to light-hole enabled at
  $\bm{k}_{\perp}\neq 0$ by a term $\propto (k_{x}^{2}- k_{y}^{2})$ in the
  $\Gamma_{8}$ Luttinger Hamiltonian, they are not eigenstate of
  $\hat{J}_{z}$.}
\bibitem[{Hex()}]{HexcC2v}
\bibinfo{note}{The coupling described by Eq.~\ref{C2v_term} contributes to
  $\delta_{1}$ through the short-range electron-hole exchange
  $-\frac{4}{3}\delta_{0}\hat{\bm{\sigma}}^{e}\cdot\hat{\bm{J}}$, by an amount
  equal to $2\beta \delta_{0}/\Delta_{h\!-\!l}^{2}$. However, the dominant
  contribution in $\delta_{1}$ is supposed to originate from the long-range
  part of the electron-hole exchange which as a result does not permit to
  establish the sign correlation with $\beta$.}
\bibitem[{InA()}]{InAsQDs}
\bibinfo{note}{For InAs QDs, the polarization eigenaxes are in most cases
  parallel to the $\langle110\rangle$ crystallographic axes leading to $\phi=\pi/4$. As a
  result $g^{h}$ gets diagonal and actually is equal to the scalar factor $3q$
  when no heavy-hole light-hole coupling is included.}
\end{thebibliography}

\end{document}